\documentclass[prl,aps, preprintnumbers, showpacs, nofootinbib,superscriptaddress,notitlepage,twocolumn]{revtex4-2}
\usepackage[normalem]{ulem}
\usepackage{epsfig}
\usepackage{amsfonts}
\usepackage{amsmath}
\usepackage{slashed}
\usepackage{graphicx}
\usepackage{mathrsfs}
\usepackage{color}
\usepackage{mathtools}
\usepackage{simpler-wick}
\allowdisplaybreaks[4]

\newcommand{\as}{\alpha_s}

\begin{document}
	
	\title{Mass renormalization group of heavy meson light-cone distribution amplitude in QCD}

	\author{Wei Wang}
	\email{wei.wang@sjtu.edu.cn}
	\affiliation{Shanghai Key Laboratory for Particle Physics and Cosmology,
		Key Laboratory for Particle Astrophysics and Cosmology (MOE),
		School of Physics and Astronomy, Shanghai Jiao Tong University, Shanghai 200240, China}
	\affiliation{Southern Center for Nuclear-Science Theory (SCNT), Institute of Modern Physics, Chinese Academy of Sciences, Huizhou 516000, Guangdong Province, China}
	
	\author{Ji Xu}
	\email{xuji\_phy@zzu.edu.cn}
	\affiliation{School of Nuclear Science and Technology, Lanzhou University, Lanzhou 730000, China}
		\affiliation{School of Physics and Microelectronics, Zhengzhou University, Zhengzhou, Henan 450001, China}
	
	\author{Qi-An Zhang}
	\email{zhangqa@buaa.edu.cn}
   \affiliation{School of Physics, Beihang University, Beijing 102206, China}
	
	\author{Shuai Zhao}
	\email{zhaos@tju.edu.cn}
	\affiliation{Department of Physics, Tianjin University, Tianjin 300350, China}
	
	\begin{abstract} 
		
The heavy meson light-cone distribution amplitude (LCDA), as defined in full QCD, plays a key role in the collinear factorization for exclusive heavy meson production and in lattice computations of the LCDA within heavy-quark effective theory (HQET). In addition to its dependence on the renormalization scale, the QCD LCDA also evolves with the heavy quark mass. We derive a partial differential equation to characterize the mass evolution of the heavy meson QCD LCDA, examining the heavy quark mass dependence through its solution. Our results link the internal structure of heavy mesons across different quark masses, offering significant implications for lattice calculations and enabling the extrapolation of results from lower to higher quark masses.

	\end{abstract}

	\date{\today}
	
	\maketitle

%\section{Introduction}
%{\it Introduction.---}

The study of heavy meson decays, such as those of the $B$-meson, offers deep insights into the nonperturbative aspects of QCD and tests the boundaries of the Standard Model. Central to these analyses is the heavy meson light-cone distribution amplitude (LCDA), which captures the internal dynamics of the meson in terms of its partonic structure. Defined through the meson-to-vacuum matrix element of light-cone-separated operators within heavy quark effective theory (HQET)~\cite{Grozin:1996pq}, the LCDA is a key element in the factorization theorems that underpin the analysis of $B$-meson exclusive decays~\cite{Beneke:1999br,Beneke:2000ry,Keum:2000wi,Lu:2000em,Cui:2023yuq} and in light-cone sum rules~\cite{Khodjamirian:2006st,Faller:2008tr,Gubernari:2018wyi,Wang:2015vgv,Wang:2017jow,Lu:2018cfc,Gao:2019lta}. As a nonperturbative quantity driven by low-energy QCD effects, the LCDA cannot be computed directly with perturbation theory. Historically, studies of the $B$-meson LCDA have relied on various phenomenological models and techniques, as detailed in Refs.~\cite{Grozin:1996pq,Korchemsky:1999qb,Beneke:1999br,Ball:2003fq,Braun:2003wx,Wang:2015vgv,Beneke:2018wjp,Gao:2021sav,Shen:2021yhe}. Significant efforts have also been made to calculate the $B$-meson LCDA directly from the principles of QCD~\cite{Kawamura:2018gqz,Wang:2019msf,Zhao:2020bsx,Xu:2022guw,Xu:2022krn,Hu:2023bba,Hu:2024ebp}. Despite these efforts, a fully model-independent determination remains elusive.

Another quantity that captures the internal structure of heavy mesons is the LCDA defined through the matrix element of the full QCD operator, referred to as the QCD LCDA~\cite{Braun:1988qv}. This amplitude accounts not only for the low-energy degrees of freedom around $\Lambda_{\mathrm{QCD}}$ but also for the region near the heavy quark mass scale, $\sim m_Q$. In collinear factorization for heavy meson production with large momentum transfer, $Q \gg m_Q$~\cite{Lepage:1980fj,Chernyak:1983ej}, the production amplitude can be factorized as a convolution of a hard scattering kernel and the QCD LCDA of the heavy meson.

The QCD LCDA also plays a pivotal role in lattice calculations of the HQET LCDA. Recently, it was proposed~\cite{Han:2024min,Han:2024yun} that the HQET LCDA for heavy mesons can be calculated model-independently by combining the large-momentum effective theory (LaMET)~\cite{Ji:2013dva} (see Refs.~\cite{Cichy:2018mum,Ji:2020ect} for reviews), HQET, and lattice simulations. In this approach, the quasi-distribution amplitude, defined with the QCD operator, is first evaluated on the lattice. It is then related to the QCD LCDA through a perturbative matching relation in LaMET. Subsequently, the QCD LCDA is matched onto the HQET LCDA using the HQET matching relation~\cite{Pilipp:2007sb,Ishaq:2019dst,Zhao:2019elu,Beneke:2023nmj}, which has been studied in both momentum and coordinate space.

The QCD LCDA depends on two scales: the renormalization scale $\mu$ and the heavy quark mass $m_Q$. Its evolution with respect to $\mu$ is governed by the Efremov-Radyushkin-Brodsky-Lepage (ERBL) equation~\cite{Efremov:1979qk,Lepage:1979zb}, similar to the evolution equation for the LCDA of light mesons. However, the QCD LCDA also exhibits dependence on the heavy quark mass, a feature that has not been systematically explored in the literature.

Understanding this heavy quark mass dependence is of particular interest, as it has significant implications for lattice calculations. Due to computational limitations, lattice simulations can only reach certain heavy quark masses, often lower than those encountered in physical scenarios. If we can precisely characterize the dependence of the LCDA on the heavy quark mass, we can extrapolate results from simulations at lower masses to predict the LCDA at higher masses that are challenging to simulate.

In this Letter, we address the following question: given the QCD LCDA at a specific heavy quark mass, $m_{Q_0}$, what can be inferred about the QCD LCDA at a different mass, $m_Q$? We show that this mass dependence can be described and predicted model-independently through the solution of a mass evolution equation.

%{\it Heavy meson LCDAs in HQET and QCD.---}
In the heavy meson exclusive decays, a frequently used LCDA is defined in the infinite-quark mass limit within HQET as
\begin{align}
	&\varphi_+ (\omega,\mu)=-i F(\mu) \nonumber\\
	\times&\int \frac{d z^-}{2\pi} e^{i \omega v^+ z^-}  \langle 0| \bar q(z)[z,0]\slashed n \gamma_5 h_v(0) | \bar H(v)\rangle,
\end{align}
where $h_v$ denotes the heavy quark field in HQET, $v$ is the velocity of meson with $v^2=1$, and $F(\mu)$ is the decay constant in HQET. The $\mu$ dependence is described by the Lange-Neubert equation~\cite{Lange:2003ff}.

Conversely, the heavy meson LCDA defined in full QCD is instrumental in lattice calculations of the $B$-meson HQET LCDA and in high-energy exclusive production of heavy mesons. In full QCD, the LCDA for a heavy meson is given by
\begin{align}
	&\phi(u,m_Q;\mu)=-i f_H \nonumber\\
	\times& \int \frac{dz^-}{2\pi} e^{i u P^+ z^-} \langle 0 | \bar q(z^-) [z^-,0]\slashed n \gamma_5 Q(0)| \bar H(P_H)\rangle,
\end{align}
where $m_Q$ is the mass of heavy quark $Q$ with $m_Q\gg \Lambda_{\mathrm{QCD}}$, $u$ is the light-cone momentum fraction of the light-quark, $P_H$ is the momentum of the heavy meson $\bar H$, $f_H$ is the decay constant, and $[z,0]$ is a Wilson line along the light-cone. $\mu$ denotes the renormalization scale. 
Notably, the QCD LCDA also depends on the heavy quark mass $m_Q$. Because $\mu\geq m_Q\gg \Lambda_{\mathrm{QCD}}$, the $\mu$ and $m_Q$ dependence of $\phi(u,m_Q;\mu)$ can be solved perturbatively.

%We study the $m_Q$ dependence with two approaches.
%{\it Mass evolution equation.---}
In the following, we derive a partial differential equation for $\phi(u, m_Q; \mu)$ that enables us to compute the dependence of the LCDA on the heavy quark mass through its solution. As noted, the heavy meson LCDA is influenced by two primary scales: the heavy quark mass $m_Q$
and the renormalization scale $\mu$. Therefore, the evolution of the heavy-meson QCD LCDA represents a two-scale problem—a common feature in QCD.

Two-scale evolution arises in other areas of QCD as well. For example, transverse-momentum-dependent parton distributions (TMDs) evolve with respect to both the renormalization scale and rapidity, with the latter described by the Collins-Soper equation~\cite{Collins:1981uk,Collins:1981va,Collins:1984kg}. Another example is the evolution of quasidistributions, where, in addition to the renormalization group equation (RGE), the quasidistribution depends on the momentum of the hadron, with momentum evolution governed by the so-called momentum RGE~\cite{Ji:2014gla,Ji:2020ect}. The renormalization scale evolution of the heavy meson LCDA follows the same equation as the RGE for light mesons, known as the ERBL equation. However, the evolution with respect to the heavy quark mass has yet to be explored in detail.

To understand the mass evolution equation, we begin with the HQET LCDA, which has ultraviolet (UV) divergences. In phenomenological analyses, the standard renormalization scheme is the $\overline{\mathrm{MS}}$ scheme, where the scale dependence follows the Lange-Neubert equation. Alternatively, one may introduce a UV cutoff such as the heavy quark mass $m_Q$, leading to a dependence of the LCDA on this cutoff 
$m_Q$. However, $m_Q$ alone is insufficient to regularize all UV divergences, requiring additional UV regularization, such as dimensional regularization (DR). Consequently, the combined evolution with respect to $m_Q$ and the $\overline{\mathrm{MS}}$ scale $\mu$ completes the scale evolution of the QCD LCDA.
In comparison, for TMDs, DR alone does not fully renormalize all UV divergences, necessitating an additional regulator like a rapidity cutoff. This two-scale regularization parallels the approach taken for the $m_Q$ evolution of the QCD LCDA.

A convenient approach to derive the equation is to ultilize HQET.
It has been demonstrated that in the infinite mass limit $m_Q\to \infty$, the QCD and HQET LCDAs can be linked with QCD perturbative theory. According to Refs.~\cite{Ishaq:2019dst,Zhao:2019elu,Beneke:2023nmj}, to the leading power of $\Lambda_{\mathrm{QCD}}/m_Q$, the QCD and HQET LCDAs can be matched by a matching relation, 
\begin{align}
		\phi(u,m_Q;\mu)=\int d\omega C(u,\omega,m_Q;\mu,\mu_F) \varphi_+(\omega,\mu_F),
\end{align}
where $\mu_F$ denotes the factorization scale, and $\mu$ is the renormalization scale for $\phi$. 
There are logarithmics of $\mu/\mu_F, \mu/m_Q$ in the matching coefficient $C(u,\omega,m_Q;\mu,\mu_F) $. To avoid the large logarithmic of $\mu/\mu_F$ one can set $\mu=\mu_F$. In this case, the factorization takes the following form~\cite{Beneke:2023nmj}
\begin{align} \label{eq:matching:peak}
\phi(u,m_Q;\mu)= \mathcal{J}(m_Q,\mu)   m_Q \varphi_+(u m_Q,\mu). 
\end{align}
This matching formula holds in the ``peak region'' $u\sim \Lambda_{\mathrm{QCD}}/m_Q$, while in the ``tail region'' $u \sim 1 $ the LCDA becomes perturbative, as discussed in Ref.~\cite{Beneke:2023nmj}.

Before proceeding, we clarify our notation for the heavy quark mass. Throughout this paper, the quark mass refers to the 
$\overline{\mathrm{MS}}$
mass, which depends on the 
$\overline{\mathrm{MS}}$
scale $\mu$. Specifically, the 
$\overline{\mathrm{MS}}$
mass $m_Q$ is defined as the mass renormalized at the scale
$\mu=m_Q$, i.e., $m_Q=m_Q(\mu=m_Q)$.

To derive the mass evolution equation,  we calculate the derivative of $\phi$ with respect to $\ln m_Q$. Using the inverse of the matching formula in  Eq.~\eqref{eq:matching:peak}, we obtain
\begin{align} \label{eq:massRGE}
	 m_Q & \frac{\partial}{\partial  m_Q}  \phi(u,m_Q;\mu)-u \frac{\partial}{\partial u} \phi(u,m_Q; \mu)\nonumber\\
	&- (1+\gamma(m_Q,\mu))	\phi(u,m_Q;\mu)=0 ,
\end{align}
where 
\begin{align}
	\gamma(m_Q,\mu)\equiv  \frac{d \ln \mathcal{J}(m_Q,\mu)}{d\ln m_Q}  .
\end{align}
We identify this equation as the mass evolution equation for the $B$-meson QCD LCDA. The dependence on $m_Q$ can be determined by solving this equation with the appropriate initial condition.

%{\it The solution of the equation.---}
To calculate the function $\gamma(m_Q,\mu)$, we first need to determine 
 $\mathcal{J}(m_Q,\mu)$, which can be obtained using  the standard matching procedure based on Eq.~\eqref{eq:matching:peak}. To avoid large logarithmic terms in $\mu/m_Q$,  the matching calculation is performed at  $\mu=m_Q$, followed by evolving  $\mathcal{J}$ from $\mu=m_Q$ to $\mu$. The evolution of $\mathcal{J}$ can be derived by matching the $\mu$ dependence of the HQET LCDA and the QCD LCDA in the peak region. The result reads
\begin{align}\label{eq:rge:coef}
	\frac{d \ln \mathcal{J}(m_Q,\mu)}{d\ln \mu}&=\frac{\alpha_s(\mu)C_F}{\pi}  \bigg( (1+\gamma_m)\ln\frac{\mu}{ m_Q}+1\bigg)\nonumber\\
	&\approx \frac{\alpha_s(\mu)C_F}{\pi}  \bigg( \ln\frac{\mu}{ m_Q}+1\bigg),
\end{align}
where $\gamma_m\equiv -d\ln m_Q(\mu)/d \ln \mu$ is the mass anomalous dimension, with the value $\gamma_m = \frac{3\alpha_s(\mu) C_F}{2\pi}  +\mathcal{O}(\alpha_s^2)$. The term associated with $\gamma_m$ is neglected because it begins at $\mathcal{O}(\alpha_s^2)$. By  calculating at $\mu=m_Q$, the matching coefficient $\mathcal{J}$ is initially determined as
\begin{align}
	\mathcal{J}(m_Q,m_Q)=1+\frac{\alpha_s(m_Q)}{4\pi} C_F \bigg(4+\frac{\pi^2}{12}\bigg)+\mathcal{O}(\alpha_s^2),
\end{align}
which provides the initial condition. This result aligns with the matching coefficient found in Ref.~\cite{Beneke:2023nmj}.
The matching coefficient is then evolved from $m_Q$ to $\mu$ by solving the renormalization group equation (RGE) in Eq.~\eqref{eq:rge:coef}, yielding
\begin{align}
	& \mathcal{J}(m_Q,\mu)=\exp\bigg[\int_{m_Q}^{\mu} \frac{d\mu'}{\mu'}\frac{\alpha_s(\mu')C_F}{\pi}  \left(\ln\frac{\mu'}{ m_Q} +1\right)\bigg] \nonumber\\
	&\quad \quad \quad \quad \quad\quad \times \mathcal{J}(m_Q, m_Q)\nonumber\\
	&\approx\exp\bigg\{\frac{ 2 C_F}{\beta_0} \bigg[ \frac{2\pi}{\beta_0} \bigg(\frac{1}{\alpha_s(\mu)}-\frac{1}{\alpha_s(m_Q)} \nonumber\\
	&\quad +\frac{1}{\alpha_s(m_Q)}\ln\frac{\alpha_s(\mu)}{\alpha_s(m_Q)}\bigg) -\ln\frac{\alpha_s(\mu)}{\alpha_s(m_Q)} \bigg]\bigg\} \nonumber\\
&\quad \times\bigg[1+	\frac{\alpha_s(m_Q) C_F}{4\pi}      \bigg(4  +\frac{\pi^2}{12}\bigg)   +\mathcal{O}(\alpha_s(m_Q)^2)\bigg] ,
\end{align}
where $\frac{d\alpha_s (\mu)}{d\ln \mu}=\beta (\alpha_s (\mu))\approx -\frac{\alpha_s^2(\mu)}{2\pi}\beta_0$ with $\beta_0 = \frac{11}{3}C_A -\frac43 n_f T_F$. 
The above result can be expanded around $\alpha_s (m_Q)$,
\begin{align}
		 \mathcal{J}&^{\mathrm{exp}}(m_Q,\mu)
		=1+\frac{\alpha_s(m_Q)C_F}{4\pi}  \nonumber\\
		&\times\left(\frac12 \ln^2 \frac{\mu^2}{m_Q^2} +2\ln\frac{\mu^2}{m_Q^2}+4  +\frac{\pi^2}{12}\right) + \mathcal{O}(\alpha_s^2),
\end{align}
which matches the result in Ref.~\cite{Beneke:2023nmj} up to $\mathcal{O}(\alpha_s^2)$.

With the above the result, we obtain the value of $\gamma(m_Q,\mu)$,
\begin{align}
	&	\gamma(m_Q,\mu)=  \frac{d \ln  \mathcal{J}(m_Q,\mu)}{d\ln m_Q} \nonumber\\
\approx & - \frac{\alpha_s(m_Q)C_F}{\pi}  -\int_{m_Q}^{\mu} \frac{d\mu'}{\mu'}\frac{\alpha_s(\mu')C_F}{\pi} \nonumber\\
&  +\beta (\alpha_s (m_Q)) \frac{C_F}{4\pi} \bigg(4  +\frac{\pi^2}{12}\bigg) .
\end{align}
Because $\beta(\alpha_s)$ starts at order $\alpha_s^2$, one can neglect the term associated with it and get
\begin{align} \label{eq:dimension}
	\gamma(m_Q,\mu)
	\approx & \frac{\alpha_s(m_Q)C_F}{\pi}-\frac{2 C_F}{\beta_0} \ln\frac{\alpha_s(\mu)}{\alpha_s (m_Q)} .
\end{align}

The mass evolution equation is a first-order partial differential equation  that can be solved using standard techniques such as the method of characteristics.  Given the initial condition
$	\phi(u,m_{Q_0},\mu)\equiv \phi_0(u)$,
the solution to the evolution equation is
\begin{align}
		\phi(u, m_Q;\mu)=& \exp\bigg[-\int_{m_{Q_0}}^{m_Q} \frac{d m'}{m'} \gamma(m',\mu)\bigg] \nonumber\\
		&\times\frac{m_Q}{m_{Q_0}}\phi_0\left(u \frac{m_Q}{m_{Q_0}}\right) .
\end{align}
With Eq.~\eqref{eq:dimension}, one can express the solution in a compact form
\begin{align}\label{eq:solution}
	&\phi(u, m_Q;\mu) 
	\approx \exp\bigg[   \frac{2  C_F}{\beta_0} \ln\frac{\alpha_s (m_Q)}{\alpha_s (m_{Q_0})} \nonumber\\
	&-\frac{4\pi  C_F}{\beta_0^2}    \bigg(  \frac{1}{\alpha_s(m_{Q_0})}\ln\frac{\alpha_s(\mu) }{\alpha_s(m_{Q_0})e}  \nonumber\\
	&-\frac{1}{\alpha_s(m_Q)}\ln\frac{\alpha_s(\mu) }{\alpha_s(m_Q)e} \bigg)   \bigg] \frac{m_Q}{m_{Q_0}}\phi_0\left(u \frac{m_Q}{m_{Q_0}}\right)  . 
\end{align}
In this solution, the LCDAs at masses $m_Q$ and $m_{Q_0}$are connected through a multiplicative factor, with the argument $u$ scaled by $m_Q/m_{Q_0}$. Consequently, as $m_Q$ increases, the LCDA becomes narrower and its peak shifts to smaller values of $u$. Additionally, the dependence on the heavy quark mass is mediated through $\alpha_s$.
To check the above solution, one can expand it around $\alpha_s(m_{Q_0})$,
\begin{align}\label{eq:solutionexp}
	&\phi^{\mathrm{exp}}(u, m_Q;\mu) 
	\approx  \bigg\{1+\frac{\alpha_s(m_{Q_0})}{2\pi}C_F\bigg[\ln^2\frac{m_Q}{m_{Q_0}}\nonumber\\
	&~~~~-2 \ln\frac{m_Q}{m_{Q_0}} \left(1+\ln\frac{\mu}{m_{Q_0}}\right)\bigg] \bigg\}\frac{m_Q}{m_{Q_0}}\phi_0\left(u \frac{m_Q}{m_{Q_0}}\right) ,
\end{align}
which matches the result calculated from the fixed order calculation. 
\begin{figure}
	\centering
	\includegraphics[width=0.42\textwidth]{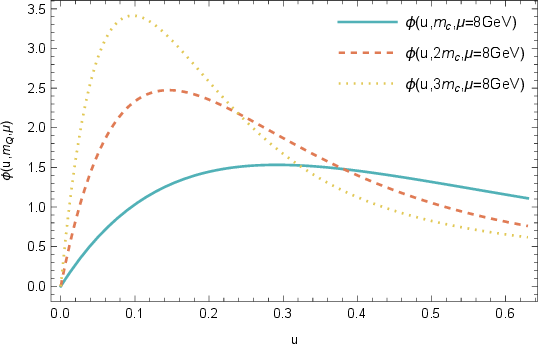}
	\caption{The evolution of heavy meson QCD LCDA with respect to heavy quark mass. The renormalization scale is chosen as $\mu=8~\mathrm{ GeV}$.  The solid line corresponds to the LCDA with $m_Q=m_c$, which works as the initial condition of the mass evolution;  the dashed and dotted lines curves correspond to $m_Q=2m_c, 3m_c$ respectively, which are calculated from Eq.~\eqref{eq:solution}.}
	\label{fig:MassEvol}
\end{figure}
\begin{figure}
	\centering
	\includegraphics[width=0.42\textwidth]{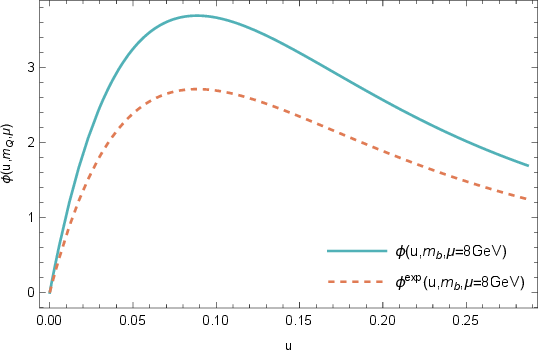}
	\caption{Comparision of the solution of mass evolution equation Eq.~\eqref{eq:solution} with its expansion around $\alpha_s(m_Q)$ (i.e., the fixed order calculation). The scale is chosen as $\mu=8~\mathrm{GeV}$. The initial condition is the QCD LCDA calculated at $m_Q=m_c$, while the curves are the QCD LCDAs evolved to $m_Q=m_b$.  }
	\label{fig:CompareExp}
\end{figure}

To illustrate our results we perform a model-based numerical calculation. For the initial condition, we adopted the QCD LCDA calculated from the matching relation Eq.~\eqref{eq:matching:peak}. We adopt the following model for HQET LCDA, in which the 
exponential model is evolved with leading-logarithmic (LL) evolution 
to the matching scale $\mu$, with analytic solution~\cite{Beneke:2018wjp,Beneke:2022msp}
\begin{align}
	\label{eq:phiexpevol}
	\varphi_+^{\rm exp}&(\omega;\mu) = e^{V+2a\gamma_E} \Gamma(2+a)\biggl( \frac{\mu_0}{\omega_0}\biggr)^a \nonumber\\
	&\times \frac{\omega}{\omega_0^2}\phantom{,}_1F_1\Bigl(2+a,2,-\frac{\omega}{\omega_0}\Bigr)\,,
\end{align}
with~\cite{Lee:2005gza}
\begin{align}
	a&= \frac{\Gamma_0}{2\beta_0}\ln r + \mathcal{O}(\as)  \,,\nonumber\\
	V&=\frac{\Gamma_0}{4\beta_0^2}\biggl[\frac{4\pi}{\as(\mu_0)}\biggl(-\ln r+1-\frac{1}{r}\biggr) \nonumber\\
	&+ \frac{\beta_1}{2\beta_0}\ln^2 r +\frac{2\gamma_0}{\Gamma_0}\beta_0 \ln r  \nonumber\\
	&+\biggl(\frac{\Gamma_1}{\Gamma_0}-\frac{\beta_1}{\beta_0} \biggr)(\ln r -r+1)\biggr] + \mathcal{O}(\as)\,,
	\label{eq:aVdef}
\end{align}
where $r=\as(\mu)/\as(\mu_0)$ and $\gamma_+(\as) =\gamma_0 \as C_F/(4\pi)+\mathcal{O}(\as^2)$ with $\gamma_0=-2C_F$, while the QCD beta function and the cusp anomalous dimensions are defined as
\begin{align}
	\label{eq:betacuspdef}
	&\beta(\as) = \mu\frac{d\as}{d\mu} = -2\as \sum_{n=0}^{\infty} \beta_n \biggl(\frac{\as}{4\pi}\biggr)^{n+1},\nonumber\\
	& \Gamma_{\rm cusp}(\as) = \sum_{n=0}^\infty \Gamma_n \biggl(\frac{\as}{4\pi}\biggr)^{n+1}\,,
\end{align}
with $\beta_0 = 11-\frac{2}{3}n_f$, $\beta_1 = 102-\frac{38}{3}n_f$ and $\Gamma_0=4C_F$, $\Gamma_1 = 4C_F(\frac{67}{3}-\pi^2-\frac{10}{9}n_f)$. We also adopt $\mu_0=1~\mathrm{GeV}$ and $\omega_0=0.3~\mathrm{GeV}$.

Fig.~\ref{fig:MassEvol} shows the mass dependence of the heavy meson QCD LCDA from Eq.~\eqref{eq:solution}. The renormalization scale is chosen as $\mu=8~\mathrm{GeV}$. We plot the LCDAs with three different values of heavy quark mass: $m_Q=m_c,~2m_c,~3m_c$, with $m_c=1.27~\mathrm{GeV}$. The LCDAs exhibits the behavior as discussed above: when $m_Q$ increases the LCDA becomes narrower and with higher value of the peak. We also compare the results from fixed order calculation~Eq.~\eqref{eq:solutionexp} with our solution Eq.~\eqref{eq:solution}, as shown in Fig.~\ref{fig:CompareExp}. The initial condition of the evolution is the LCDA with $m_Q=m_c$ and $\mu=8~\mathrm{GeV}$, and then the LCDA is evolved to $m_b=4.18~\mathrm{GeV}$.
One can read from the figure that the mass evolution modifies the fixed order calculation significantly.

At last we add some remarks on the tail region. According to the tail region matching calculation in Ref.~\cite{Beneke:2023nmj}, the tree-level diagram does not contribute, the leading contribution comes from one-loop diagrams. Therefore, to derive the mass evolution equation for the tail region, one should calculate two-loop diagrams. We leave it for a future study.

%{\it Conclusions.---} 
To summarize, we have derived a mass evolution equation for the heavy meson QCD LCDA in the ``peak'' region, which holds significant phenomenological importance. By solving this equation, we determine how the QCD LCDA depends on the heavy quark mass, relating the LCDA of heavy mesons with different quark masses. Together with the ERBL equation, our results complete the two-scale evolution of the 
heavy meson LCDA.
Moreover, our findings facilitate the extrapolation of lattice calculations of the heavy meson QCD LCDA performed at smaller heavy quark masses to those with larger quark masses. Our results can be validated by comparing them with future lattice calculations of the heavy meson QCD LCDA.

\section*{Acknowledgements} 

We thank Yan-Bing Wei for the helpful discussions. This work was supported in part by National Natural Science Foundation of China under grant Nos. 12125503,  12335003, 12475098, 12105247 and 12375069. This work was partially supported by SJTU Kunpeng\&Ascend Center of Excellence. Q. A. Z. was also supported by the Fundamental Research Funds for the Central Universities.

%%%%%%%%%%%%%%%

\end{document}